\documentclass[letterpaper]{sfchem}
\usepackage{graphicx}

\newcommand{\ump}{$\mu$m}
\newcommand{\um}{$\mu$m }

\newcommand{\hi}{{\sc Hi} }

\begin{document}

\title{Depletion of small dust grains in the early stages of molecule formation}

\author{ M.-A. Miville-Desch\^enes\inst{1}}
  \institute{Canadian Institute for Theoretical Astrophysics (CITA), Toronto, Canada}
%Short author list here: Surnames only please (no initials)
\authorrunning{Miville-Desch\^enes}
%Short title here:
\titlerunning{}

\maketitle 

\begin{abstract}
ISOCAM allowed the observation of diffuse
interstellar regions where the transition from atomic to molecular gas takes place.
In this contribution we report on spectacular variations of small dust grains abundance in such regions
observed with ISOCAM. In the four cirrus-like clouds observed, we note a systematic disappearance 
of small dust grains in regions where CO emission starts to appear. 
We suggest that the variations of the small dust grain abundance observed here 
are related to the damping of the gas turbulent motions that favors the coagulation
of dust through grain-grain collisions. 

\keywords{interstellar medium; dust; turbulence}

\end{abstract}

\section{Introduction}

We know that interstellar dust plays an important role in the physics  
of the interstellar medium. For instance, dust grains heat 
the gas via the photo-electric effect and provide the surfaces necessary to the
formation of H$_2$. These processes are sensitive to the size distribution, the 
structure and composition of dust grains and, especially, to the abundance of the 
smallest particles. The IRAS images of nearby molecular complexes 
and bright cirrus clouds ($A_V > 1$) have revealed remarkably that
the abundance ratio between small and large dust grains widely vary from cloud to cloud 
and within clouds.
The gain in brightness sensitivity and angular resolution provided since then by dedicated observations
carried out with ISOCAM, the camera on board the Infrared Space Observatory (ISO), 
now allows us to carry on these IRAS investigations and in particular to extend them
to the range of  column densities ($N_{\rm H} \sim 5\times10^{20}$ cm$^{-2}$) where the transition
from atomic to molecular gas takes place.\\
It is the motivation of this work to shed further light
on the evolution of small dust grains within the diffuse interstellar medium
in relation to gas physical conditions. 

In this contribution we present mid-IR imaging observations of four
high Galactic latitude cirrus obtained with ISOCAM at 6" angular resolution.
The observations were done with two filters LW2 (5-8.5 \ump) and LW3 (12-18 \ump),
that measure respectively the aromatic carbon bands and the underlying 
continuum emission from small dust particles.
The comparison of these observations with 21 cm, CO and IRAS data  allows us
to measure the mid-IR emissivity per hydrogen which, in such
clouds transparent to stellar light, is related to the abundance and the optical properties of 
small dust particles independently of any modeling of the 
penetration of the dust heating radiation. 

\section{New results on infrared cirrus}

The ISOCAM database contains several observations of high-latitude interstellar 
regions that can be used to study the evolution of dust in the diffuse interstellar medium. 
In this section, I present ISOCAM observations of four diffuse clouds
that confirm the very fast evolution of small dust grains in the early stages of 
molecular gas formation.

%%%%%%%%%%%%%%%%%%%%%%%%%%%
%%     URSA MAJOR
%%%%%%%%%%%%%%%%%%%%%%%%%%%
\subsection{Ursa Major}

The first diffuse interstellar region we have studied was the Ursa Major cirrus. 
The details of this analysis are described in Miville-Desch\^enes et al. (2002).
We have observed three 0.05 square degree fields in the Ursa Major
Galactic cirrus with ISOCAM in the LW2 and LW3 filters. These observations
of a weakly illuminated and diffuse atomic cloud revealed
striking variations of the properties of small dust particles, responsible
for the emission in the 4-18 \um range. First, the combination
of ISOCAM images and 21 cm interferometric data allows us to
highlight a rotating filament with a small dust particles abundance
$\sim 5$ times higher than its surrounding. Second, an unexpected variation of 
the mid-infrared color was found, at an interface between a fully atomic
region and a denser part of the cloud where molecular hydrogen is formed
and survives (see Fig~1). The comparison of the ISOCAM observations
with IRAS data suggests a spectacular decrease of the small
dust grain abundance in the CO region, suggestive of grain coagulation.

%%%%%%%%%%%%%%%%%%%%%%%%%%%
%%     POLARIS 
%%%%%%%%%%%%%%%%%%%%%%%%%%%
\subsection{Polaris}

Fig.~2 presents a region of the Polaris flare observed 
at 100 \um (IRAS), 21 cm (DRAO) and in the mid-infrared (ISOCAM).
This region of the Polaris flare shows very different structure at IRAS 
100 \um and in the 21 cm integrated emission map.
In particular, the brightest filament observed at 100 \um is not seen
at all at 21 cm, indicating that this diffuse filamentary structure is mainly composed of molecular gas.
What is most interesting here is that this filament does not show any evidence of mid-infrared
emission attributed to PAH-like particles (ISOCAM observations - see Fig.~2 top-right). 
In fact, in the restricted region observed with ISOCAM, there is a very clear spatial correlation between the 
integrated 21 cm and the mid-infrared emissions, indicating that most of the mid-infrared emission comes from the
atomic gas and that the abundance of small dust grains in the molecular gas is considerably decreased.

%%%%%%%%%%%%%%%%%%%%%%%%%%%
%%     L1780
%%%%%%%%%%%%%%%%%%%%%%%%%%%
\subsection{L1780}

Based on IRAS observations Laureijs et al. (1991) showed that the central part of the
L1780 cloud, which has a moderate opacity (A$_v <4$), is characterized by a low 
60/100 \um ratio. This has been interpreted as the sign of a low dust temperature (T$\sim 15$ K)
and a change in the Very Small Grains (VSG) properties.  
The mid-infrared observations of the L1780 cloud done with ISOCAM show that
the decrease of the 6.5/100 \um ratio in the central part of the filament is even more spectacular
(see Fig.~3).
These observations are a strong indication that the mid-infrared emitters (pah-like particles)
are rapidly coagulating on bigger grains in the denser part of the cloud, where CO emission is detected.
This scenario is in accordance with the hypothesis of  Laureijs et al. (1991) who state that the
change in the VSGs optical properties may be due to the formation of dirty-ice mantle.

%%%%%%%%%%%%%%%%%%%%%%%%%%%
%%     TAUFIL
%%%%%%%%%%%%%%%%%%%%%%%%%%%
\subsection{Taurus}

Non star-forming regions with low 60/100 \um ratio values 
(often referred as {\bf cold} infrared emission) are generally associated
with CO emission. A tight correlation between these two quantities
has been remarkably shown in the Taurus complex by Abergel et al. (1994). 
We extend this analysis with mid-infrared ISOCAM observations of a particular filament in this complex (see Fig.~4).
Like in the three other regions observed in this study, the Taurus filament observed with ISOCAM has very
low 6.5 \um emission associated with a low 60/100 \um and CO emission. 
Again, these observations are in accordance with a spectacular decrease of the
abundance of the mid-infrared emitters in diffuse molecular gas. On the other hand, 
as the optical opacity are moderate in this Taurus filament, like in the L1780 filament, we cannot completely rule out the effect of absorption.

\section{Discussion}

Thanks to ISOCAM's high sensitivity and to the data processing techniques that allow us
to reach it (Miville-Desch\^enes et al. 2000) we are now able to study the properties
of very small grains even in diffuse cirrus clouds.
From the comparison with gas phase observations (\hi and CO), 
such observations bring a strong evidence that interstellar turbulence has an 
impact on the dust size distribution. As dust grains are coupled to the gas motions, 
turbulent motions of great amplitude may produce grain-grain collision strong enough to fragment
big grains and to enhance the abundance of small grains. On the other hand,
when turbulent motions are damped, like in the molecular gas, grain-grain collisions may lead to
the disappearance of small grains through coagulation. 

The ISOCAM observations of such regions also raise interesting questions regarding the
evolution of the interstellar medium in general. As most of the reactive surface for the photo-electric effect and
for the formation of molecules is on small grains, the strong variations of small dust grains abundance
reported here suggest significant spatial variations of the chemical and thermal 
balance in the diffuse ISM. In particular, the rapid disappearance of small dust grains
in molecular regions raises the question of the H$_2$ formation rate in such environment.
In addition, if small dust grains coagulate that rapidly, we can wonder if there are
any of them in condensed regions. It also raises the question of the origin of small
dust grains at the surface of molecular clouds and in PDRs.

\begin{figure*}[!ht]
\includegraphics[width=\linewidth, draft=false]{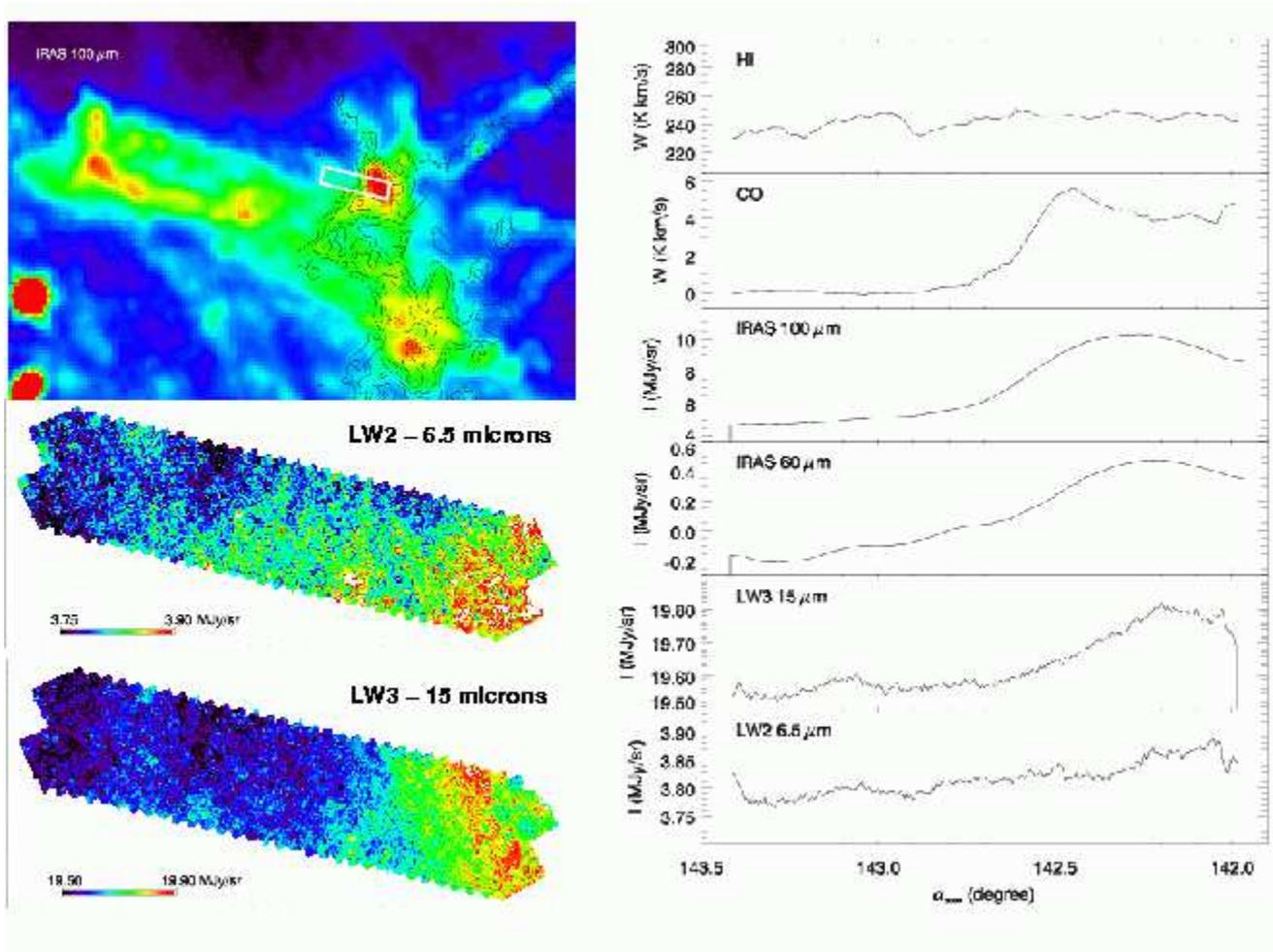}
\caption{{\bf Left: } IRAS 100 \um (top) emission of the Ursa Major cirrus (contours are the $^{13}$CO 
integrated emission of Pound \& Goodman (1997)) and 6.5 \um (middle) and 15.0 \um (bottom) ISOCAM observations
of the HI-H$_2$ interface (shown as a white rectangle). {\bf Right: } Average emission
as a function of right ascension through the HI-H$_2$ interface.
Note that the mid-infrared emission gets significantly flatter than the 100 \um emission in the molecular region,
indicating a decrease by a factor 3 of the small dust grain abundance with respect to the big grains.}
\end{figure*}

\begin{figure*}[!ht]
\includegraphics[width=\linewidth, draft=false]{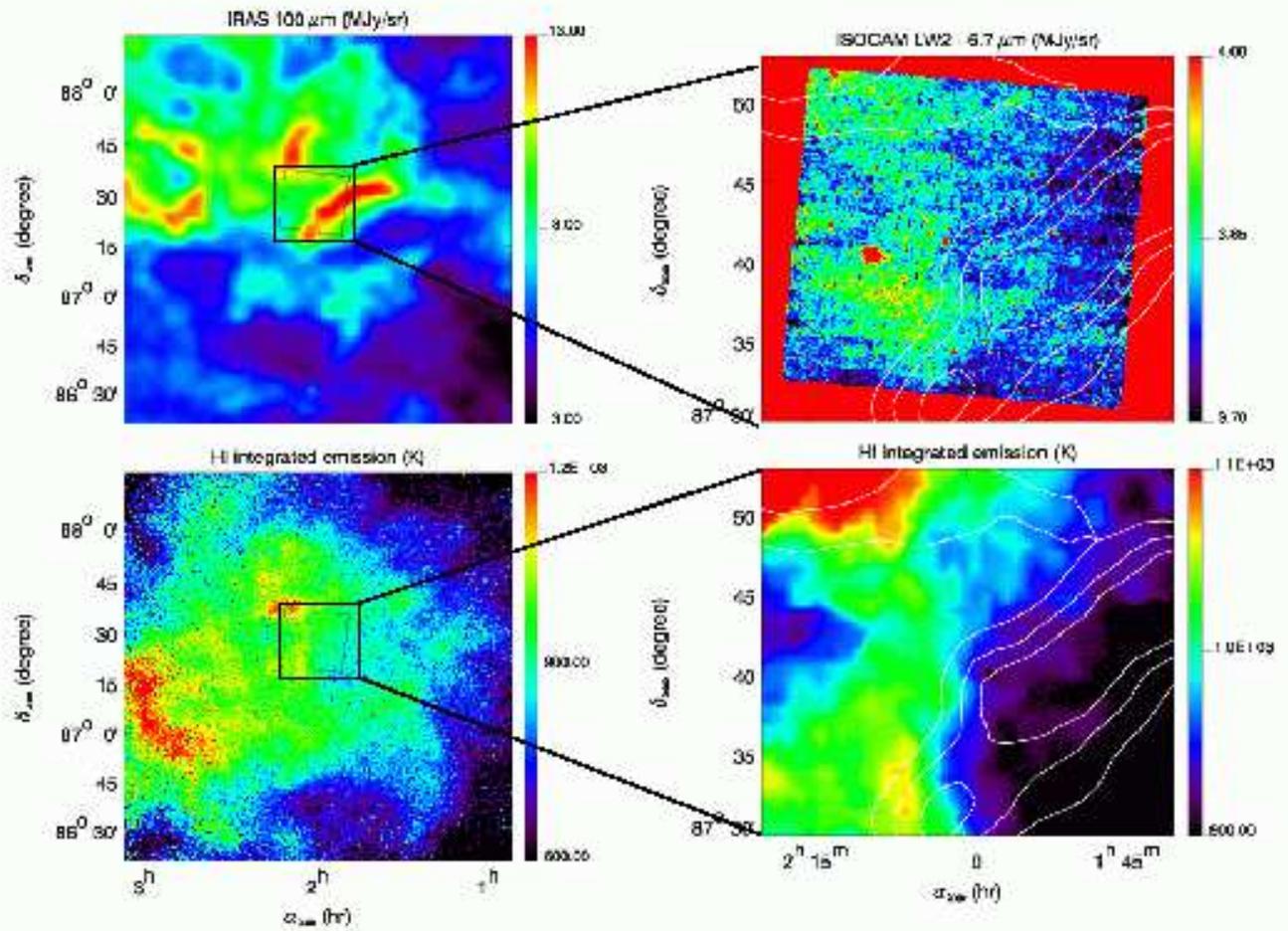}
\caption{The Polaris flare. {\bf Left: } IRAS 100 \um (top) and 21 cm integrated emission (bottom) 
images of the Polaris cirrus cloud. 
The 21 cm observations were obtained
at the Dominion Radio Astrophysical Observatory (DRAO) in Penticton. 
{\bf Right: } A smaller region has been
observed with ISOCAM at 6.5 \um (top). An enlargement of the same region seen at 21 cm is also shown (bottom).
On these last two images, the IRAS 100 \um emission is shown as contours.}
\end{figure*}

\begin{figure*}[!ht]
\includegraphics[width=\linewidth, draft=false]{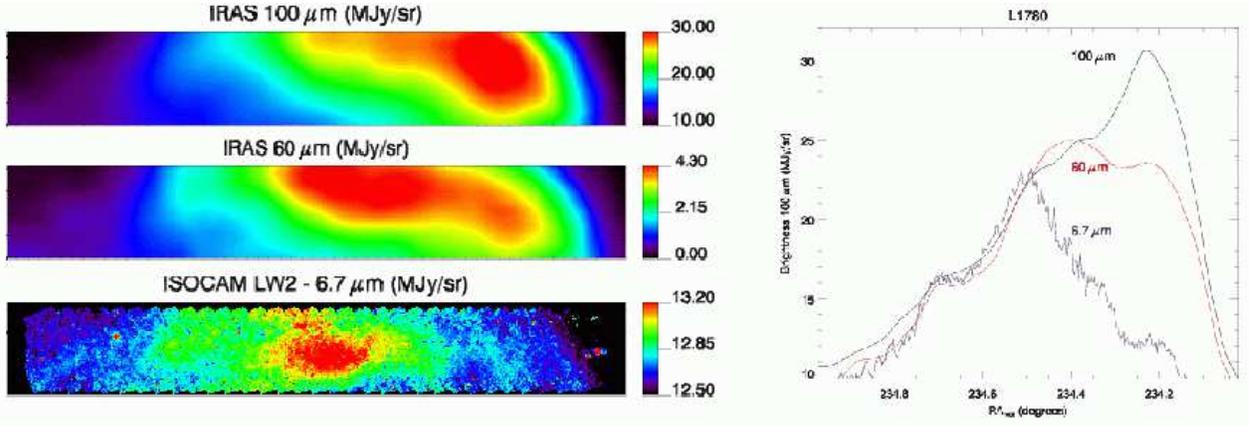}
\caption{The L1780 cloud. {\bf Left:} IRAS 100 (top) and 60 (middle) \ump, 
and ISOCAM 6.5 \um (bottom) of a cut across the L1780 cloud.
{\bf Right:} Average emission of the three tracers as a function of right ascension. 
The 6.5 and 60 \um were multiplied by standard factors to be compared to the 100 \um emission.
There is a spectacular decrease of the 6.5 / 100 \um ratio inside the L1780 filament. 
This is in accordance with a decrease of the 
smallest dust grain abundance, spatially correlated with the apparition of CO emission.}
\end{figure*}

\begin{figure*}
\includegraphics[width=\linewidth, draft=false]{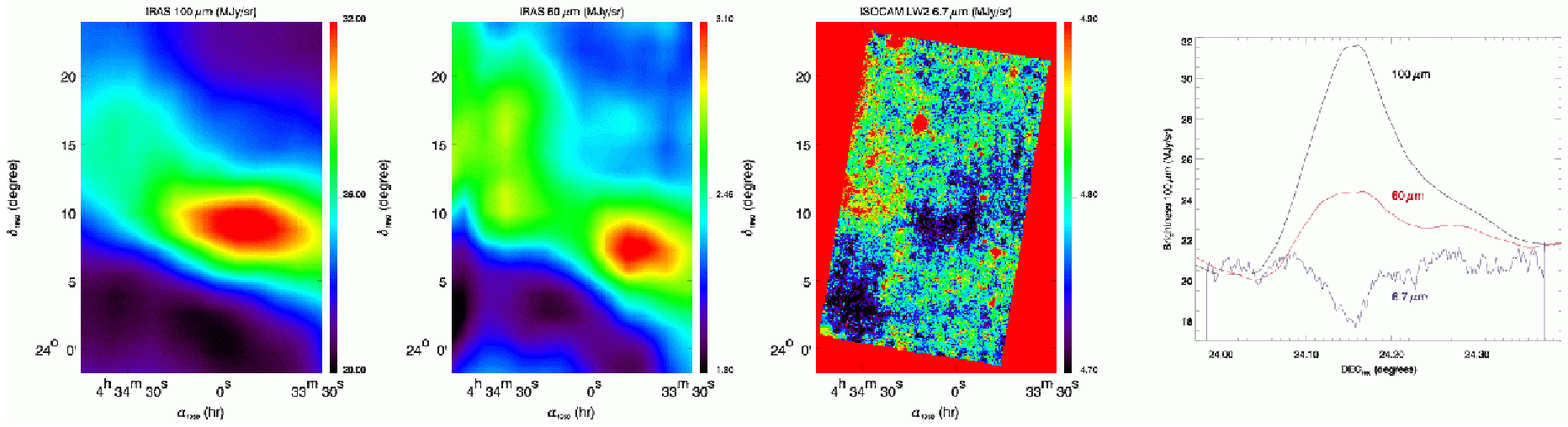}
\caption{The Taurus filament. IRAS 100 \ump, IRAS 60 \um and ISOCAM 6.5 \um observations of a filament in Taurus.
On the right, we show a cut across the filament (average emission of the three tracers as a function of declination).}
\end{figure*}


\begin{thebibliography}{}

\bibitem[]{} Abergel, A., et al.,  1994, ApJ, 423, L59

\bibitem[]{} Laureijs, R. J., Clark, F. O., Prusti, T., 1991, ApJ, 372, 185

\bibitem[]{} Miville-Desch\^enes, M.-A., et al., 2002, A\&A, 381, 209

\bibitem[]{} Miville-Desch\^enes, M.-A., et al.,  2000, A\&AS, 146, 519 

\bibitem[]{} Pound, M. W., Goodman, A. A., 1997, ApJ, 482, 334

\end{thebibliography}
\end{document}